\documentstyle[prd,aps,epsf,epsfig]{revtex} %c2
\begin{document}
\draft
\twocolumn[\hsize\textwidth\columnwidth\hsize\csname@twocolumnfalse\endcsname%c2
%:::::::::::::::::::::::::::::::::::::::::::::::::::::::::::::::::::::::
%
\title{
PHYSICS OF THE RHYTHMIC APPLAUSE}
%:::::::::::::::::::::::::::::::::::::::::::::::::::::::::::::::::::::::
\author
{Z. N\'eda, E. Ravasz}
\address{Babe\c{s}-Bolyai University, Dept. of Theoretical Physics\\
str. Kog\u{a}lniceanu nr.1, RO-3400, Cluj-Napoca, Romania}
\author
{T. Vicsek}
\address{E\"otv\"os-Lor\'and University, Dept. of Biological Physics \\
Budapest, Hungary}
\author
{Y. Brechet}
\address{LTPCM-ENSEEG/INPG, Domaine Universitaire de Grenoble\\
B.P. 75, 38402 Saint Martin d'Heres, Cedex, France}
\author
{A.L. Barab\'asi}
\address{University of Notre Dame, Department of Physics\\
IN 46556, U.S.A}
\maketitle
\centerline{\small (Last revised: 7 February 2000)}

\begin{abstract}
We report on a series of measurements aimed to characterize the
development and the dynamics of the rhythmic applause in concert
halls. Our results demonstrate that while this process shares many
characteristics of other systems that are known to synchronize, it
also has features that are unexpected and unaccounted for in many
other systems. In particular, we find that the mechanism lying at
the heart of the synchronization process is the period doubling
of the clapping rhythm. The characteristic interplay between synchronized
and unsynchronized regimes during the applause is the result of
a frustration in the systems. All results are understandable in the
framework of the Kuramoto model.
\end{abstract}


%{\bf Keyword:} rhythmic applause, Kuramoto model, synchronization,
%frustration
\vspace{2pc}
]%c2

\vspace{1cm}

%:::::::::::::::::::::::::::::::::::::::::::::::::::::::::::::::::::::::
% Text begins:
%:::::::::::::::::::::::::::::::::::::::::::::::::::::::::::::::::::::::

\section{Introduction}

Everybody experienced after a good performance the audience showing it's 
appreciation with a thunderous synchronized clapping. This is suddenly 
arising from the initially incoherent clapping of individuals and might
disappear and reappear again several times during the applause.
This synchronization process in the concert hall offers a wonderful example
of social self-organization, and it is believed to be the human
scale example of the synchronization processes observed in numerous
systems in nature. As detailed in 
\cite{strogartz} spontaneous synchronization occurs in many biological and 
sociological systems: fireflies in Southeast Asia synchronize their flashes;
crickets synchronize their chirping; neural cells of the brain synchronize
voltage fluctuations; pacemaker cells in the heart synchronize their
fire; womens living together for long times find their menstrual cycle
synchronized.

The accepted way of modeling the synchronization process is by considering
coupled non-linear oscillators.
Understanding the synchronization process of coupled oscillators
is an old problem in physics, mathematics and theoretical biology
\cite{glass}.
It is believed that the problem goes back to
Huygens, who realized first the 
synchronization of pendulum clocks hanging together on a wall.
While the synchronization of identical coupled oscillators interacting with 
phase-minimizing interactions is obvious, the problem is non-trivial
if one considers a population of non-identical oscillators. Depending
on the strength and type of the interaction, and the dispersion of the
oscillators frequencies, synchronization might, or might not appear.
For biological populations it is crucial to consider non-identical internal
frequencies, since individuals are not rigorously
equivalent. In this way in order to understand synchronization 
in biological or sociological systems we have to consider the more complex
problem of coupled non-identical oscillators.

For a theoretical study of coupled oscillator models
many interaction types were studied. Considering mathematically coupled
maps \cite{just} the problem became interesting  mostly from the view-point of
dynamical systems and chaos. Since most of the
coupling processes in nature are through pulse-like interactions
(firing of neurons, flashes of fireflies, clapping etc.)    
a very realistic pulse-like coupling  is considered in
the integrate and fire type models \cite{mirollo,bottani}.
In these models the
firing of an oscillator results in a phase-jump of all other
oscillators. Because the phase-evolution of each oscillator
is considered non-linear in time, under very general conditions 
the jumping process leads to synchronization.
Interactions continuously present in time and thus memory effects
are considered in the models introduced
by  Winfree \cite{winfree} and Kuramoto \cite{kuramoto}.
In these later models we
have phase-difference minimizing interactions between globally
coupled rotators. For a suitable chosen interaction form, the
problem becomes exactly solvable and leads to very interesting results.

Although at a first choice the integrate and fire type models
look the most promising to understand the
synchronization and dynamics of the rhythmic applause, we will
argue in favor of the phase-coupled Kuramoto model. 
According to our interpretation of the phenomenon in the appearance of
rhythmic applause memory effects are crucial.
By considering the pulse-like interaction of integrate and fire models
this important ingredient is totally neglected. In contrast,
the continuous phase-coupling in the Kuramoto model offers
a first approach to deal with the relevant memory effects.

The purpose of our present paper is to reveal
some interesting and new
peculiarities of the rhythmic applause, and to understand 
the observed phenomenon  
in the framework of the well-known 
Kuramoto model. Of course, in the context of the rhythmic applause 
the Kuramoto model should be viewed as an abstract theoretical 
representation of the situation, containing the minimum number of
assumptions already allowing the kind of synchronization exhibited 
by the spectators.
We mention that the main results of our study was briefly discussed
in a very short recent communication \cite{nature}. Here we will consider
a much more detailed and argumented presentation.

\section{The Kuramoto model}

In order to construct a simple physical model for synchronization one can
take a population of non-identical rotators globally coupled 
through phase-difference minimizing interactions.
Synchronization
in this system is not obvious at all and triggered the interesting 
studies of Winfree \cite{winfree}. He found that mutual synchronization 
is possible or not depending on the relation between the oscillators 
frequency distribution width (dispersion) and the strength of coupling.
Later, Kuramoto proposed, and together with Nishikava  exactly 
solved \cite{kuramoto} an elegant reformulation of Winfree's model. 

In the Kuramoto model we have $N$ rotators each of them described by it's
$\phi_K$ phase. The rotators  have a $g(\omega)$ distribution of
their $\omega_K$ natural frequencies. Every rotator interacts with all 
the other ones via  phase-difference minimizing terms:
\begin{equation}
W^{int}_K=\frac{K}{N}\sum_{j=1}^{N} \sin{(\phi_j-\phi_K)}
\end{equation}
The $N$ coupled differential equations describing the 
over-damped oscillators dynamics are:
\begin{equation}
\frac{d\phi_K}{dt}=\omega_K+\frac{K}{N} \sum_{j=1}^N \sin{(\phi_j-\phi_k)}
\end{equation}
Mathematically the synchronization level will be characterized by an order 
parameter, $q$, defined at any time-moment as:
\begin{equation}
q=\left| \frac{1}{N} \sum_{j=1}^N e^{i \phi_K} \right|
\end{equation}
The maximal possible value $q=1$ corresponds to total synchronization,
the case $0<q<1$ to partial synchronization, while for $q=0$ there is
no synchronization at all in the system.

In the $N\rightarrow \infty$ thermodynamic limit of the equilibrium
dynamics ($t\rightarrow \infty$, so initial transient effects are lost) 
Kuramoto and Nishikava proved the existence of a 
$K_c$ critical coupling. For a
Gaussian distribution of the oscillators natural frequencies, 
characterized by a $D$  dispersion they got \cite{kuramoto}:
\begin{equation}
K_c=\sqrt{\frac{2}{\pi^3}} D
\end{equation}
For $K\le K_c$ the only possible solution gives $q=0$ (no synchronization)
while for $K>K_c$ a stable solution with $q\ne 0$ appears.
Thus, the main result is that for a population of globally coupled 
non-identical oscillators a partial synchronization of the phases is 
possible whenever the interaction among oscillators exceeds a critical
value. By adopting the Kuramoto model for biological populations the
observed synchronization phenomenon are conceptually understood.

\section{Rhythmic applause versus the Kuramoto model: open questions}  

As we argued in the introduction it is convenient to use the Kuramoto 
model in order to describe the dynamics
of the rhythmic applause. If the Kuramoto model is applicable to the
phenomenon there could be two cases:
i.) the average coupling between the spectators is smaller than the
critical one and no synchronization appears,
ii.) for larger coupling than the critical one synchronization would gradually
evolve. The critical coupling (as discussed in section 2) is governed by the 
dispersion of the spectators natural clapping frequencies (4).

Studying carefully the rhythmic applause one
might raise immediately many questions which are not obviously answered 
within a simple application of the Kuramoto model:
\begin{itemize}
\item Usually at the beginning of the applause there is a long 
"waiting" time without any synchronization, and with no increase in the 
order parameter. The partial synchronization is evolving after that 
suddenly and achieves its maximal value in a short time. 
This should be not the case if we are in the $K>K_c$ limit. One would expect
in this limit a continuous increase in the order parameter right
from the beginning of the applause.
\item Why synchronization already achieved is lost after a time, and why
might reappear again? Loosing synchronization should not happen
in the $K>K_c$ limit.
\end{itemize}

In order to understand more deeply the phenomenon and to answer the questions
which are not obvious within the framework of the Kuramoto model we first
considered an experimental study.

\section{Experimental study}

Two main experiments were considered:

I. The applause after many good theater and opera performances
(in Romania and Hungary) was recorded,
digitized and analyzed in several aspects. Recordings both by a 
microphone hanging form the ceiling of the concert hall  
and in the neighborhood of randomly selected individuals were considered.

II. Well-controlled clapping experiments were carried out on 
a group of $73$ high-school students. We also investigated the 
clapping frequencies of one individual during a one week period.

\subsection{Experimental method}

{\bf I.} By digitizing the signal (Fig.~1a) we obtained a recording of
fluctuating voltage with both positive and negative values. 
The zero level (as the mean-signal
level) was determined and the square of the signal relative to this level
was computed (Fig.~1b). This signal would roughly correspond to the noise
intensity variation, but due to the short sampling time it is definitely
not the one that our human perception can follow. 
The average on a relatively short-time period ($\approx 0.2 s$) was 
considered and we got the signal (Fig.~1c) which describes well the useful 
noise-intensity variations during the applause. Already at a first-look one
can easily detect the part where the rhythmic applause  appears and  
the periodic
noise intensity variation during synchronized clapping.

This short-time averaged signal was analyzed in several aspects:
\begin{enumerate}
\item A long-time averaged noise-intensity was computed by averaging
the signal on a time interval of approximately $3s$.
\item An experimentally computable order-parameter, $q_{exp}$ was 
calculated. This order-parameter was defined in a very analogous way 
with the $q$ order parameter (3) in the Kuramoto model. At each time-step
$q_{exp}$ is calculated as the maximum of the normalized correlation between 
the $s(t)$ signal and a harmonic function
\begin{equation} 
q_{exp}(t)=\mbox{max}_{\{T,\phi\}} \left\{ \frac{\int_{t-T}^{t+T} s(t) 
\sin{(2 \pi/T+\phi)} dt }{\int_{t-T}^{t+T} s(t) dt } \right\}
\label{qexp}
\end{equation}
(in the above formula the values of $\phi$ should span all
possible initial phases between $0$ and $2 \pi$, and $T$ should vary 
between two reasonably chosen limiting values, $T_{min}$ and $T_{max}$.)
Taking $T_{min}=0.1s$ and $T_{max}=5s$  $q_{exp}$
was numerically computed.
\item Finally, for recordings taken in the neighborhood of one spectator 
one can observe that the short-time averaged noise intensity curves have some
evident local maximums corresponding to the clapping of the individual.
In this case we computed the time-interval between the clearly 
distinguishable groups of maxima as a function of time. This will
characterize the clapping frequency of the chosen spectator.
\end{enumerate}

{\bf II.} In our well-controlled clapping experiments
we studied the distribution function for the natural clapping
frequency of non-interacting students. Separated by the rest of the
group, students were first encouraged to clap in the manner they
would right after a good performance (mode I clapping).
Second, they were asked to clap in
the manner they would do it during synchronized clapping (mode II clapping).
On one student we also performed investigations on
the
both two ways of clapping, sampling $100$
times during a one-week period.

\subsection{\bf Experimental Results}

{\bf I.} A characteristic part of the global applause where the 
synchronization is formed and destroyed is presented as a
function of time on Fig.~2a.
On Fig.~2b. we present simultaneously the recording made in the
vicinity of one spectator. The order parameter variation computed
by equation \ref{qexp} from the global signal is plotted on Fig.~2c.
On Fig.~2d we show the variation of the long time moving average
(window size: 3s) for the global noise intensity. Finally, on
Fig.~2e we present the computed clapping period as a function of
time for the chosen individual.
One can notice immediately that during the rhythmic applause
(central part) the long-time averaged noise intensity has a clear minimum,
the order parameter has a maximum (as expected) and the
clapping period of an individual presents also a maximum.
These results are stable and qualitatively consistent with all the $47$
recordings we have studied. 

{\bf II.} For the normalized  distribution of the natural clapping 
frequencies of the group of $73$ students the results are given on Fig.~3.  
With continuous line we plotted the distribution obtained for mode I clapping  
and with dashed line
the results for mode II clapping. Both distributions are roughly Gaussian.
The maximum of the first distribution is located at a higher
frequency (approximatively two times higher) and the peak
is wider than the one for mode II clapping. Calculating the
dispersion for the two distributions, we got 
both the dispersion ($D$) and the relative dispersion
($D_r=D/\overline{\omega}$)
for mode II. clapping smaller:
\begin{eqnarray}
\frac{D_I}{D_{II}} \approx 2.5 \\
\frac{D^r_{I}}{D^r_{II}} \approx 1.3
\end{eqnarray}

Computing the normalized distribution for
the ratio of mode I to mode II clapping for each
individual we got also a one-peek
distribution centered around the value of $2$ (Fig.~4). 

The normalized distribution of mode I and mode II clapping obtained
from $100$ measurements on one individual (Fig.~5) shows a similar
behavior with the one presented in Fig.~4.
We plotted with continuous line the normalized distribution
obtained for mode I clapping and with dashed line the one for mode II
clapping. The distribution of possible natural frequencies during
mode II clapping is again sharper than the one for mode I
clapping. Again, the location of the
maxima for the distribution of mode I clapping is roughly double than
the one for mode II clapping.

\section{Discussion}

From the controlled clapping experiments   
(Figs.~3-5) we conclude that spectators can clap with two
very distinct clapping modes. During rhythmic applause
they shift from their original high frequency mode I clapping
to the slow frequency, mode II clapping.
The above picture is totally 
supported by the clapping rhythm of one individual during the
rhythmic applause (Fig.~2e). 
During this period-doubling  process
the dispersion (width) of the natural clapping frequencies distribution
is decreased roughly to half. This is a very important step
in understanding the phenomenon from the viewpoint of the Kuramoto model.
From the Kuramoto model we learned that the critical coupling, $K_c$,
for synchronization is directly proportional with the dispersion of the
frequency distribution of the oscillators (4). When the dispersion is 
decreased the value of the critical coupling decreases too.
When the clapping starts the
spectators clap very enthusiastically and with high frequencies (mode I
clapping). The frequency distribution is wide and the value of the
critical coupling is large. The coupling between the spectators is
lower than the critical value and synchronization is not possible.
The spectators inconstiently realize this (or already inconstiently
know the "game" of the rhythmic applause)
and by roughly doubling their clapping periods
they shift to mode II clapping. By this the dispersion of the natural
frequencies of the spectators reduces to half and so does also the value
of the $K_c$ critical coupling.
The coupling among spectators can become in this way larger
than $K_c$ and partial synchronization (rhythmic applause) appears.

It is clear now why synchronization appears, but the question remains:
why after a time this synchronization is lost?
The answer is immediate 
by examining the long-time averaged noise intensity curve (Fig.~2c).
One can realize that
after synchronization occurs the average noise decreases and attains its 
minimal value. 
Enthusiastic spectators are not satisfied with this and try to
increase the average noise intensity level. This is possible mainly by
speeding up the rhythm, because the intensity of one clap cannot be 
increased over a given level any more simply by hitting stronger.
Speeding up the rhythm increases the natural frequency distribution's
dispersion and consequently increases by
this the value of $K_c$. The coupling among spectators becomes again
smaller than the critical one and synchronization is lost.

The above scenario is fully supported by the analyzed signal 
in Fig.~2. We learn from here that during the rhythmic applause
the individuals in the crowd double their clapping period (Fig.~2e),
and we observe how by this the
order-parameter  characterizing synchronization is increased (Fig.~2d).
When the order-parameter is maximum the average noise intensity (Fig.~2c)
is minimum and spectators will increase this by increasing
the frequency of clapping shifting again to mode II clapping.

The clapping after a good performance is frustrated in
some sense. When maximal synchronization is achieved the
average noise intensity is minimum due to the slow frequency of
mode II clapping.
By increasing the clapping frequency and shifting to mode I clapping
the value of critical coupling is increased and synchronization is lost.
The spectators cannot achieve both maximal noise intensity and good
synchronization within the same clapping mode, and this makes
the system frustrated.

A very simple computer simulation exercise on the Kuramoto model can 
visualize all the above results (Fig.~6). Considering an ensemble of
$N=70$ globally coupled rotators with a Gaussian distribution of their 
natural frequencies ($\overline{\omega}=2\pi s^{-1}$, $D=2 \pi/6.9 s^{-1}$ 
and $K=0.8 s^{-1}$) we follow up
their dynamics. We associate a very small time length $\tau=0.01s$ and 
$I_0=\omega/\overline{\omega}$ intensity noise for each event 
when a rotator passes through 
an integer multiple of phase $2\pi$. The global noise is obtained by simply 
adding all noises generated by the individual rotators. 
In the given setup we have $K<K_c$, and thus no
synchronization should appear. However, at $t_1=21 s$ time moment we double the 
natural oscillation period  of each oscillators, and as observable in 
Fig.~6a synchronization gradually evolves. This leads to an increase 
in the order-parameter (Fig.~6b, computed now as in experiments after
equation \ref{qexp}), and a decrease in the long-time  averaged noise
intensity (Fig.~6c). Beginning with $t_2=35s$ we begin to linearly increase the
rotators natural frequencies back to their original values. As observable,
this will decrease again the order parameter and increase the averaged noise 
intensity. This computer exercise approximates well the experimental  results
from Fig.~2, and gives confidence in our analytical statements.

Some interesting comments can be made at this point.
\begin{enumerate}
\item In many societies the phenomenon of rhythmic applause is unknown.
It seems that the "game" of the rhythmic applause has to be learned
by each community. Spectators once inconstiently recognizing that
lowering the clapping frequency allows synchronization and
getting familiar to the phenomenon they will keep alive this
habit.
\item Usually, in huge open air concerts no clear rhythmic applause
forms even after outstanding performances.
This is mainly due to the small and
non-global coupling existing in the system (the applause of far-away
spectators is totally undetectable). Even by
reducing to half the natural clapping frequencies dispersion it
is still not possible to achieve a coupling larger than the critical one.
\item In communist times it was a common habit to applaud by
rhythmic applause the "great" leader speech. During this rhythmic applause
the synchronization was almost never lost.
This is a very nice evidence of the fact that spectators
were not enthusiastic enough and were satisfied with the obtained
global noise intensity level, having no desire to increase it.
Frustration was not present in this system.
\end{enumerate}

\section{Conclusions}     

By recognizing two characteristic 
clapping modes of the individuals during the applause and by
applying the results of the Kuramoto model the dynamics of the
rhythmic applause is qualitatively understood.
The rhythmic applause is formed by lowering
(roughly to half) the natural clapping frequency of each individual,
and thus by lowering the width of the clapping frequency distribution
of the audience.        
The rhythmic applause is lost and might reaper again due
to two main desires of
the spectators which can not be both fulfilled at the same time: 
{\bf optimal synchronization} and {\bf maximal applause intensity}.
The system is frustrated in this sense. When maximal synchronization
is achieved the average noise intensity is minimum due to the slow
clapping frequency. Increasing the average noise intensity
by increasing the clapping frequency synchronization gets lost.
The amazing interplay between unsynchronized
and synchronized clapping is a consequence of this frustration.

%:::::::::::::::::::::::::::::::::::::::::::::::::::::::::::::::::::::
% References:
%:::::::::::::::::::::::::::::::::::::::::::::::::::::::::::::::::::::::

%:::::::::::::::::::::::::::::::::::::::::::::::::::::::::::::::::::::::
% Figures Captions:
%:::::::::::::::::::::::::::::::::::::::::::::::::::::::::::::::::::::::

\onecolumn

%...........................
% Fig.1
%...........................

\begin{figure}[htp]
\epsfig{figure=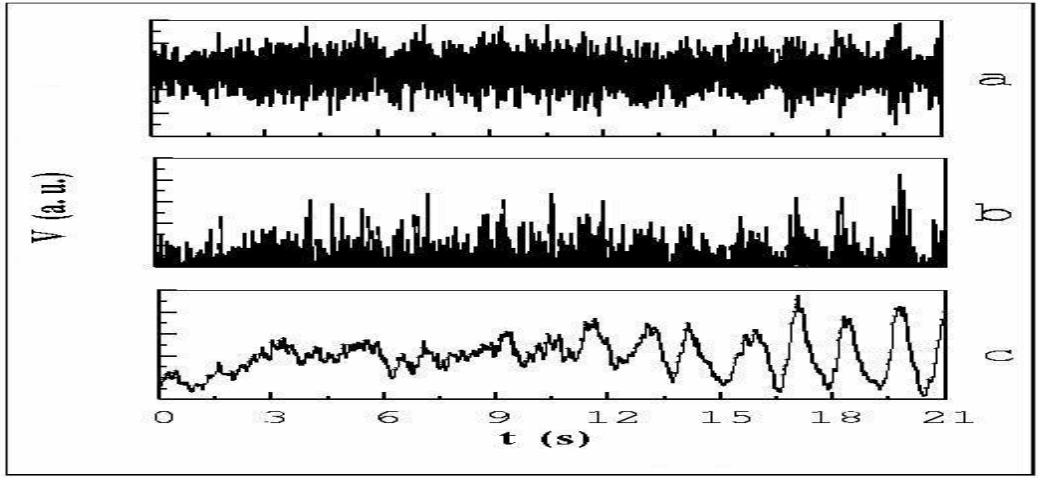,height=3in,width=6in,angle=-0}
\caption{Digitized global signal,
voltage as a function of time ($a$). The processed signals:
($b$) and ($c$). On figure $b$ we have the square of the
signal from figure $a$, relative to the calculated average voltage level.
On figure $c$ we present the short-time moving
average (window size: $0.2s$) of the signal from figure $b$.}
\label{fig1}
\end{figure}

%.............................
% Fig. 2
%.............................

\begin{figure}[htp]
\epsfig{figure=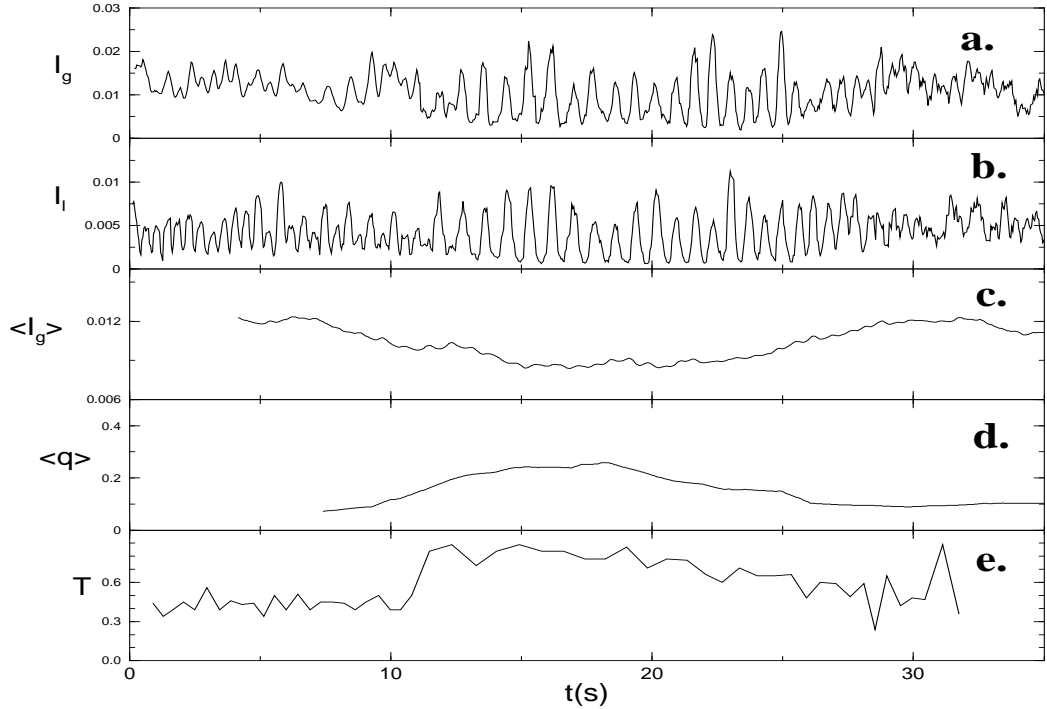,height=4in,width=6in,angle=-0}
\caption{ Short-time averaged global ($a$) and local ($b$) signal
as a function of time.
The computed long-time averaged signal (window size: $3s$) is
presented on figure $c$, the experimental order parameter
on figure $d$, and the clapping period of the chosen individual
on figure $e$. 
}
\label{fig2}
\end{figure}

%.............................
% Fig. 3
%.............................

\begin{figure}[htp]
\epsfig{figure=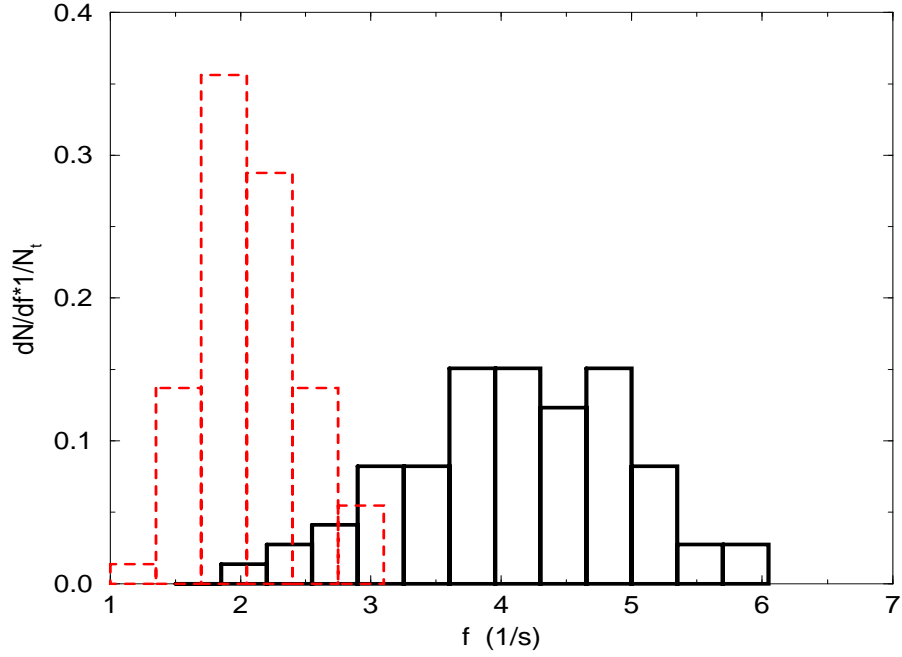,height=4in,width=5.5in,angle=-0}
\caption{ Normalized distribution functions for mode I (continuous line)
and mode II (dashed line) clapping frequencies on a
sampling of $73$ high-school students.
}
\label{fig3}
\end{figure}

%.............................
% Fig. 4
%.............................

\begin{figure}[htp]
\epsfig{figure=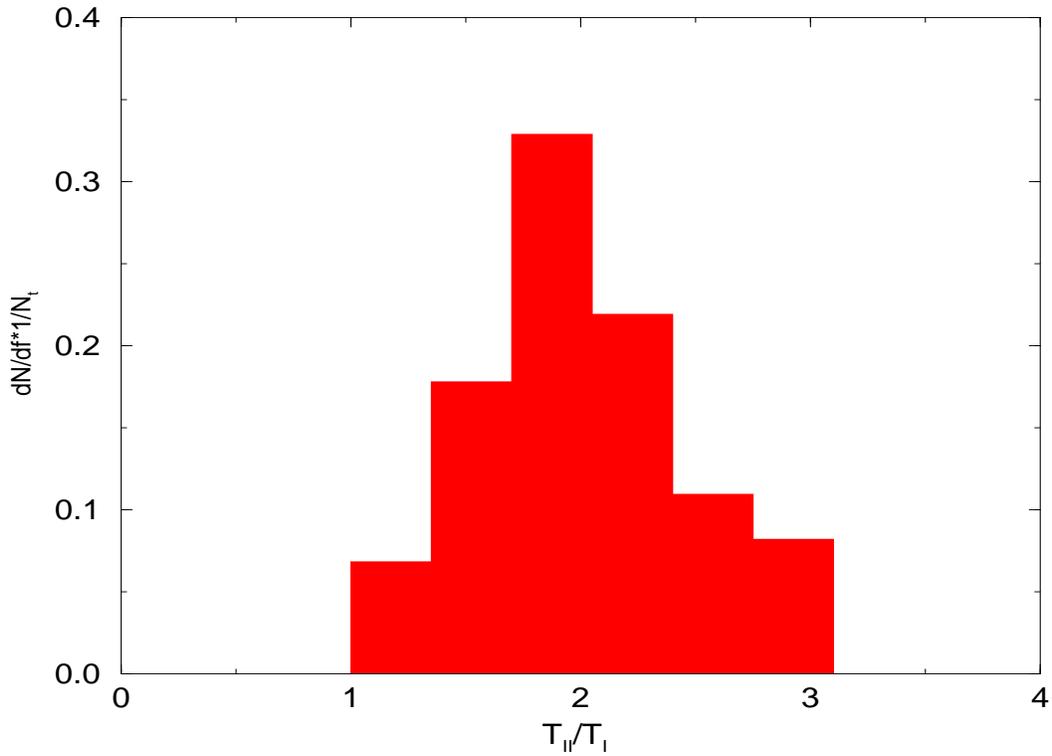,height=4in,width=5.5in,angle=-0}
\caption{Normalized distribution functions for the ratio of
the frequencies of mode I  and mode II clapping
(sampling on $73$ high-school students).
}
\label{fig4}
\end{figure}

%.............................
% Fig. 5
%.............................

\begin{figure}[htp]
\epsfig{figure=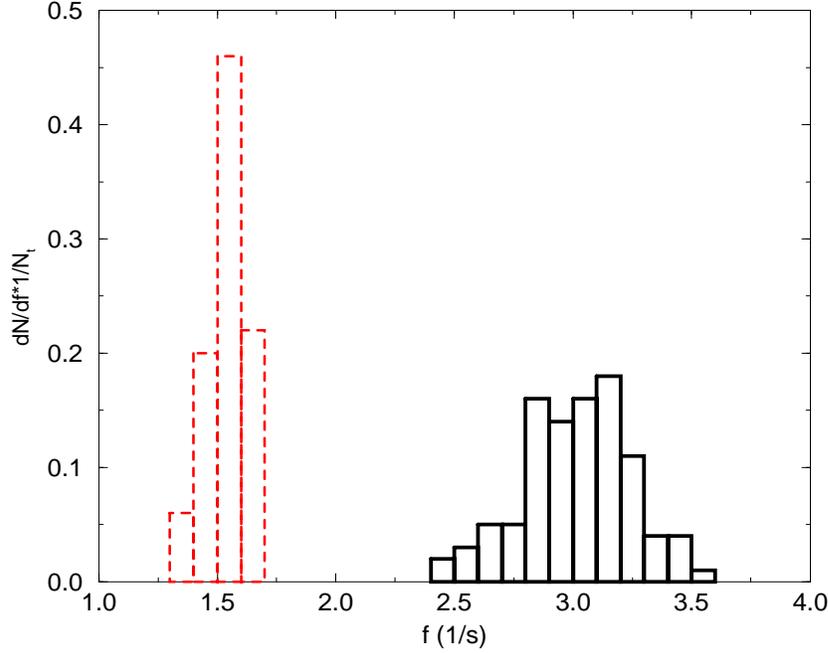,height=4in,width=5in,angle=-0}
\caption{Normalized distribution functions for the frequencies of
mode I (continuous line) and mode II (dashed line) clapping of one chosen
individual (sampling 100 times during a one week period).
}
\label{fig5}
\end{figure}

%.............................
% Fig. 6
%.............................

\begin{figure}[htp]
\epsfig{figure=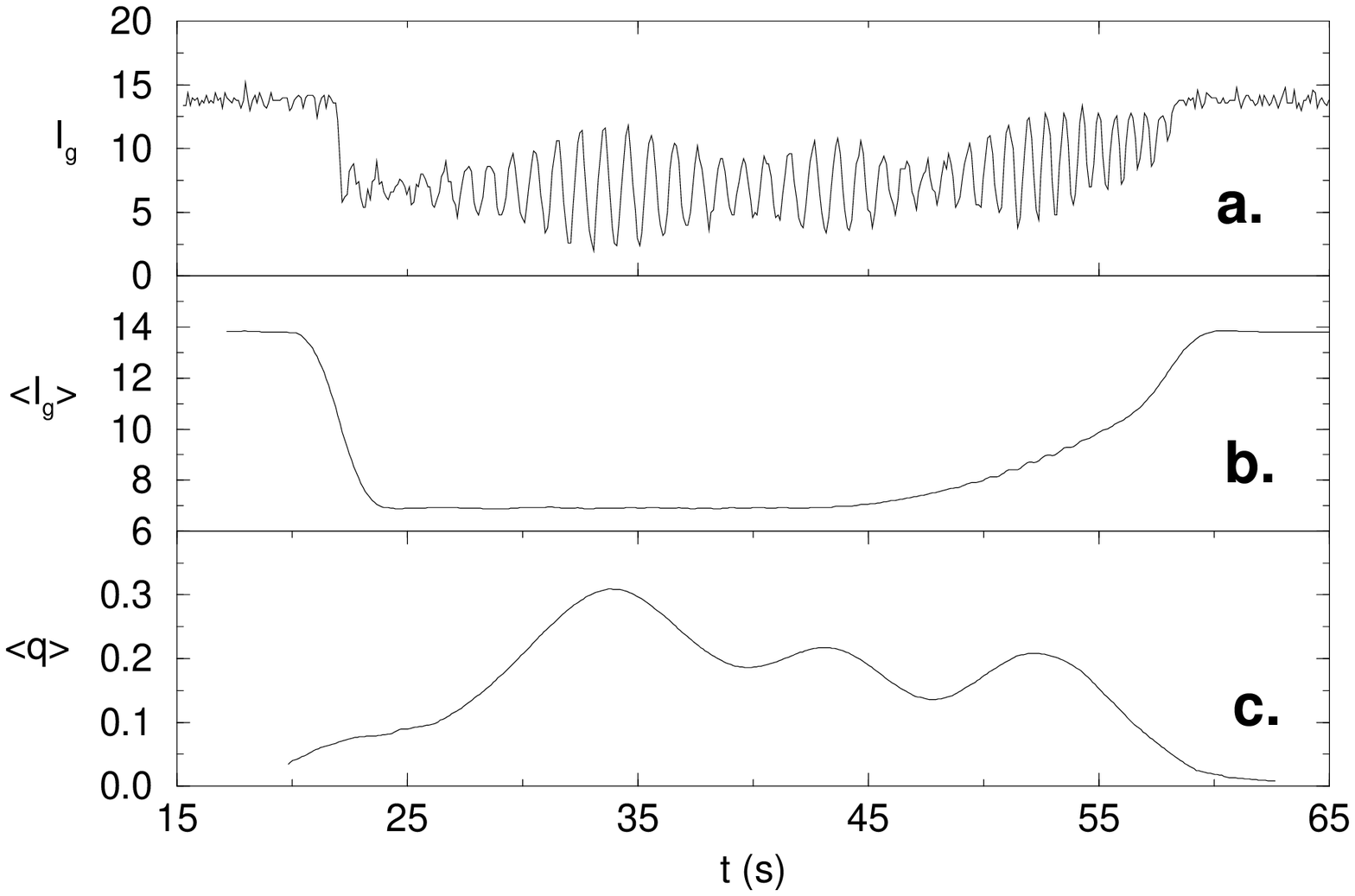,height=4in,width=5in,angle=-0}
\caption{Computer simulation of the Kuramoto model for $N=70$ rotators
($K=0.8 s^{-1}$, $\overline{\omega}=2 \pi s^{-1}$ and $D=2 \pi/6.9 s^{-1}$). 
We double the rotators
periods at $t_1=21s$ and linearly increase the frequency back to the 
original value from $t_2=35s$. The noise-pulse given by one oscillator has 
$I_0=\omega/\overline{\omega}$ intensity and $\tau=0.01s$ duration.
}
\label{fig6}
\end{figure}


\begin{references}

\bibitem{strogartz}
S.H. Strogatz and I. Stewart, {\em Scientific American}, December 1993,
p. 102;
S.H. Strogatz, {\em Lecture Notes in Biomathematics} {\bf 100} (1993)
\bibitem{glass}
L. Glass and M. Mackey, {\em From Clocks to chaos: The Rhythms of Life}
(Princeton University Press, Princeton, N.J, 1988)
\bibitem{just}
W.Just, {\em Phys. Rep.} {\bf 290}, 101 (1997); Y. Jiang, {\em
Phys. Rev. E} {\bf 56}, 2672 (1997); M.G. Rosenblum, A.S. Pikovsky and
J. Kurths, {\em Phys. Rev. Lett.} {\bf 76}, 1804 (1996)
\bibitem{mirollo}
R. Mirollo and S. Strogatz, {\em SIAM J. Appl. Math.} {\bf 50}, 1645 (1990)
\bibitem{bottani}
S. Bottani, {\em Phys.Rev. E} {\bf 54}, 2334 (1997)
\bibitem{winfree}
A.T. Winfree; {\em J. Theor. Biol.} {\bf 16}, 15 (1967)
\bibitem{kuramoto}
Y. Kuramoto and I. Nishikava; {\em J. Stat. Phys.} {\bf 49}, 569 (1987)  
\bibitem{nature}
Z. N\'eda, E. Ravasz, Y. Brechet, T. Vicsek and A.L. Barab\'asi; 
Nature (in press)


\end{references}
\end{document}